%Paper: hep-ph/9309325
%From: FTNIR@WEIZMANN.WEIZMANN.AC.IL
%Date: Sun, 26 Sep 93 11:57:46 +0200

\input phyzzx
\input tables
%macropackage=phyzzx
\def\({[}
\def\){]}
\def\ra{\rightarrow}
\def\plotpicture#1#2#3{\vskip#2}

\REF{\bigi}{I.I. Bigi {\it et al.}, in {\it CP Violation},
 ed. C. Jarlskog (Singapore, World Scientific, 1989).}
\REF{\niqu}{Y. Nir and H.R. Quinn,
 Ann. Rev. Nucl. Part. Sci. 42 (1992) 211.}
\REF{\ssi}{Y. Nir, in the Proc. of the 20th SLAC Summer Institute
 on Particle Physics, SLAC-REPORT-412, ed. L. Vassilian (1992).}
\REF{\dln}{C.O. Dib, D. London and Y. Nir,
 Int. J. Mod. Phys. A6 (1991) 1253.}
\REF{\gronau}{M. Gronau, these proceedings.}
\REF{\geho}{J.-M. Gerard and W.-S. Hou, Phys. Rev. D43 (1991) 2909.}
\REF{\siwy}{H. Simma and D. Wyler, Phys. Lett. B272 (1991) 395.}
\REF{\soar}{J.M. Soares, Nucl. Phys. B367 (1991) 575.}
\REF{\biwa}{I.I. Bigi and S. Wakaizumi, Phys. Lett. B188 (1987) 501.}
\REF{\tss}{M. Tanimoto, Y. Suetaka and K. Senba,
 Z. Phys. C40 (1988) 539.}
\REF{\wht}{S. Wakaizumi, T. Hayashi and M. Tanimoto, Phys. Rev. D39
 (1989) 2792.}
\REF{\hasu}{T. Hasuike {\it et al.}, Int. J. Mod. Phys. A4 (1989) 2465;
 Phys. Rev. D41 (1990) 1691.}
\REF{\waka}{S. Wakaizumi {\it et al.},
 Phys. Rev. D44 (1991) 3582.}
\REF{\lond}{D. London, Phys. Lett. B234 (1990) 354.}
\REF{\nisib}{Y. Nir and D. Silverman, Phys. Rev. D42 (1990) 1477.}
\REF{\silv}{D. Silverman, Phys. Rev. D45 (1992) 1800.}
\REF{\bpmr}{G.C. Branco, P.A. Parada, T. Morozumi and M.N. Rebelo,
 Phys. Rev. D48 (1993) 1167.}
\REF{\ecgr}{G. Ecker and W. Grimus, Z. Phys. C30 (1986) 293.}
\REF{\lowy}{D. London and D. Wyler, Phys. Lett. B232 (1989) 503.}
\REF{\liwo}{J. Liu and L. Wolfenstein, Phys. Lett. B197 (1987) 536.}
\REF{\dogo}{J.F. Donoghue and E. Golowich, Phys. Rev. D37 (1988) 2542.}
\REF{\gena}{J.-M. Gerard and T. Nakada, Phys. Lett. B261 (1991) 474.}
\REF{\sowoa}{J.M. Soares and L. Wolfenstein, Phys. Rev. D46 (1992) 256.}
\REF{\grni}{Y. Grossman and Y. Nir, Phys. Lett. B313 (1993) 126.}
\REF{\sowo}{J.M. Soares and L. Wolfenstein, Phys. Rev. D47 (1993) 1021.}
\REF{\hawe}{L. Hall and S. Weinberg, Phys. Rev. D48 (1993) 979.}
\REF{\biga}{I.I. Bigi and F. Gabbiani, Nucl. Phys. B352 (1991) 309.}
\REF{\lnsb}{M. Leurer, Y. Nir and N. Seiberg, RU-93-43 (1993).}
\REF{\nisa}{Y. Nir and U. Sarid, Phys. Rev. D47 (1993) 2818.}
\REF{\hara}{L. Hall and A. Rasin, LBL-33668 (1993).}
\REF{\niqub}{Y. Nir and H.R. Quinn, Phys. Rev. D42 (1990) 1473.}
\REF{\tani}{M. Tanimoto, Phys. Rev. Lett. 62 (1989) 2797.}
\REF{\thss}{M. Tanimoto, K. Hirayama, T. Shinmoto and K. Senba,
 Z. Phys. C48 (1990) 99.}
\REF{\coma}{D. Cocolicchio and L. Maiani, Phys. Lett. B291 (1992) 155.}
\REF{\bpmra}{G.C. Branco, P.A. Parada, T. Morozumi and M.N. Rebelo,
 Phys. Lett. B306 (1993) 398.}
\REF{\nise}{Y. Nir and N. Seiberg, Phys. Lett. B309 (1993) 337.}
\REF{\lns}{M. Leurer, Y. Nir and N. Seiberg,
 Nucl. Phys. B398 (1993) 319.}
\REF{\nisia}{Y. Nir and D. Silverman, Nucl. Phys. B345 (1990) 301.}

%\nopagenumbers

\rightline{WIS-93/94/Aug-PH}
\centerline{{\bf NEW PHYSICS EFFECTS ON CP VIOLATION IN $B$ DECAYS}
\foot{Plenary talk presented at the Workshop on B Physics
at Hadron Accelerators, Snowmass, Colorado, June 21 -- July 2, 1993.}}
\vskip 1cm

\centerline{YOSEF NIR}
\centerline{\em Physics Department, Weizmann Institute of Science,}
\centerline{\em Rehovot 76100, Israel}
\vskip 1cm

\leftline{\bf 1.\ \ \ INTRODUCTION}
We review new physics effects on CP violation in $B$ decays.
For previous reviews on this subject, we refer the reader to refs.
$\(\bigi,\niqu,\ssi,\dln\)$.
A discussion of CP violation in $B$ decays within the Standard Model
(and a guide to the literature) can be found in $\(\gronau\)$.

In chapter 2 we introduce our formalism, and discuss the
Standard Model picture of CP violation in $B$ decays, with
special emphasis on the cleanliness of the predictions.
Chapter 3 gives a general discussion of new physics effects:
we point out the ingredients in the analysis that are sensitive
to new physics and deduce the type of new physics that is most
likely to modify the Standard Model predictions.
Explicit examples are given in chapter 4: a model with
$Z$-mediated flavor changing neutral currents (FCNC) demonstrates
in which ways will new physics manifest itself in CP asymmetries
in $B$ decays; a supersymmetric model with ``quark--squark alignment"
mechanism shows that supersymmetry may affect CP asymmetries in
$B$ decays, even though the minimal supersymmetric Standard Model (MSSM)
does not; multi-scalar models may affect the asymmetries even in
the absence of new CP violating phases; schemes for quark mass
matrices will be crucially tested by the CP asymmetries.
In chapter 5 we explain how, if deviations from the Standard Model
predictions are measured, we will be able to learn detailed features
of the New Physics that is responsible for that.
\endpage

\leftline{\bf 2.\ \ \ THEORETICAL BACKGROUND}
Let us first describe our basic formalism.
A more detailed discussion can be found in ref. $\(\ssi\)$.
If $B$ and $\bar B$ are the CP conjugate bottom mesons ({\it i.e.}
$B^0$ and $\bar B^0$, $B^+$ and $B^-$, $B_s$ and $\bar B_s$),
and $f$ and $\bar f$ are CP conjugate final states, then we denote
by $A$ and $\bar A$ the two CP conjugate amplitudes:
$$A\equiv\bra{f}H\ket{B},\ \ \
\bar A\equiv\bra{\bar f}H\ket{\bar B}.\eqn\aabar$$
For the neutral $B$ mesons, we define $p$ and $q$ to be the
components of the interaction eigenstates $B^0$ and $\bar B^0$
within the mass eigenstates $B_H$ and $B_L$ ($H$ and $L$ stand for
Heavy and Light, respectively):
$$\ket{B_L}=p\ket{B^0}+q\ket{\bar B^0},\ \ \
\ket{B_H}=p\ket{B^0}-q\ket{\bar B^0}.\eqn\pandq$$
For final CP eigenstates $f_{CP}$, we define the product
$$\lambda\ \equiv\ {q\over p}\ {\bar A_{f_{CP}}\over A_{f_{CP}}}.
\eqn\lamdef$$
The quantities $|\bar A/A|$, $|q/p|$ and $\lambda$ are free of
phase conventions and physical.
\par
We distinguish three types of CP violation in meson decays:
\par $(i)$ CP violation in decay:
$$|\bar A/A|\neq1.\eqn\cpdecay$$
Here, CP violation arises from the interference between direct decay
amplitudes. CP violation of the type \cpdecay\ can be observed
in non-leptonic charged $B$ decays, {\it e.g.} a difference in the
rate of $B^+\ra K^+\pi^0$ and $B^-\ra K^-\pi^0$.
\par $(ii)$ CP violation in mixing:
$$|q/p|\neq1.\eqn\cpmix$$
Here, CP violation arises from the mass eigenstates being different from
the CP eigenstates. CP violation of the type \cpmix\ can be observed
in semi-leptonic neutral $B$ decays, {\it e.g.} a difference in the
rate of $\bar B^0_{\rm phys}(t)\ra\ell^+\nu X$ and
$B^0_{\rm phys}(t)\ra\ell^-\nu X$.
\par $(iii)$ CP violation in the interference of mixing and decay:
$${\rm Im}\lambda\neq0,\ \ \ |\lambda|=1.\eqn\cpmixdec$$
Here, CP violation arises from the interference between the direct decay,
$B^0\ra f_{CP}$, and the ``first - mix, then - decay" process,
$B^0\ra\bar B^0\ra f_{CP}$.
Of course, $|\lambda|\neq1$ also reflects CP violation, but it
belongs to either or both of the types \cpdecay\ and \cpmix.
CP violation of the type \cpmixdec\ can be observed
in neutral $B$ decays into final CP eigenstates that are dominated by
a single weak phase, {\it e.g.} a difference in the
rate of $\bar B^0_{\rm phys}(t)\ra\psi K_S$ and
$B^0_{\rm phys}(t)\ra\psi K_S$.

There is a significant difference in the cleanliness of the
theoretical calculations in the three types of CP violation.
If a certain decay gets contributions from various amplitudes
with absolute values $A_i$, strong phases $\delta_i$ and weak,
CP violating phases $\phi_i$, then
$$\left|{\bar A\over A}\right|=\left|{
\sum_i A_ie^{i\delta_i}e^{-i\phi_i}\over
\sum_i A_ie^{i\delta_i}e^{+i\phi_i}}\right|.\eqn\theoi$$
It follows that direct CP violation requires both non-trivial
strong phase difference ($\delta_i-\delta_j\neq0$) and non-trivial
weak phase difference ($\phi_i-\phi_j\neq0$). Conversely, the
calculation of direct CP violation requires knowledge of strong phase
shifts and absolute values of various amplitudes and, therefore,
necessarily involves hadronic uncertainties.
\par
In the neutral $B$ system, where the width difference between the
two mass eigenstates is much smaller than the mass difference,
$$\left|{q\over p}\right|=1-{1\over2}{\rm Im}{\Gamma_{12}\over
M_{12}}.\eqn\theoii$$
While $M_{12}$ is measured by the mass difference, $\Gamma_{12}$
needs to be theoretically calculated. This is basically a long-distance
physics calculation, and therefore involves large hadronic uncertainties.
While it is clear that
$|q/p|-1$ is very small (${\cal O}(10^{-3})$), the actual value
is uncertain by a factor of a few $\(\bigi\)$.

In contrast, CP asymmetries of the type \cpmixdec\ are theoretically
clean. Take, for example, the $B\ra\psi K_S$ mode. The deviation of
$|\lambda|$ from unity due to CP violation in mixing is, as mentioned
in the previous paragraph, of order $10^{-3}$. The deviation of
$|\lambda|$ from unity due to direct CP violation is even smaller:
not only is the penguin diagram much smaller than the tree diagram,
it also carries to a good approximation the same weak phase.
Thus, the interpretation of the measured CP asymmetry in terms of
electroweak parameters, $a_{CP}(B\ra\psi K_S)=\sin2\beta$, is
accurate to better than $10^{-3}$. In other modes,
where the penguin contribution differs in phase
from the tree diagrams, hadronic uncertainties are larger,
{\it e.g.} of order 10\% in $B\ra\pi\pi$.

The Standard Model predictions for direct CP violation in
various semi-inclusive $B^\pm$ decays are given in Table 1.
We take the results for the purely hadronic modes
from refs. $\(\geho,\siwy\)$. The results
in these two references agree, except for the modes marked with
a star, where $\(\geho\)$ quotes very small asymmetries. The quoted
values should be taken as representative numbers and not as
exact predictions. The asymmetries
in the radiative decays were calculated in ref. $\(\soar\)$.
\vskip 1cm
\begintable
\multispan{3}\tstrut \ 1. Direct CP Violation \crthick
 Decay | BR | $a_{CP}$ \crthick
 $\bar b\ra\bar uu\bar s$ | $5\times10^{-3}$ | $0.006^*$ \cr
 $\bar b\ra\bar dd\bar s$ | $3\times10^{-3}$ | $0.005$ \cr
 $\bar b\ra\bar ss\bar s$ | $3\times10^{-3}$ | $0.005$ \cr
 $\bar b\ra\bar uu\bar d$ | $8\times10^{-3}$ | $-0.004^*$ \cr
 $\bar b\ra\bar ss\bar d$ | $3\times10^{-4}$ | $-0.04$ \cr
 $\bar b\ra\bar dd\bar d$ | $3\times10^{-4}$ | $-0.04$ \cr
 $\bar b\ra\bar s\gamma$  | $3\times10^{-4}$ | $0.005$ \cr
 $\bar b\ra\bar d\gamma$  | $1\times10^{-5}$ | $0.1$
 \endtable
\vskip 1cm
It is difficult, however, to see how these inclusive asymmetries can be
experimentally measured. It is more likely that direct CP violation
would be measured in exclusive modes. On the one hand side,
the asymmetries for exclusive modes could be much larger. On the other
hand, their calculation suffers from larger hadronic
uncertainties and is sometimes very sensitive to the value of $q^2$
being used. Examples of exclusive asymmetries are $\(\geho,\siwy\)$
$$\eqalign{a_{CP}(B^+\ra K^+\pi^0)&\sim0.01,\cr
a_{CP}(B^+\ra K^+K^{*0})&\sim0.05.\cr}\eqn\excsm$$
Again, the Standard Model prediction is uncertain by at least
a factor of a few in either direction. However,
if the measured asymmetries are very large, say $\gg0.2$, it would
be very difficult to accommodate them in the Standard Model
even if one stretches the hadronic uncertainties, and would
probably signal new physics.

An estimate of the Standard Model value of the CP asymmetry
in semi-leptonic $B$ decays,
$$a_{SL}\equiv{\Gamma(B^0_{phys}(t)\ra\ell^-\nu X)-
\Gamma(\bar B^0_{phys}(t)\ra\ell^+\nu X)\over
\Gamma(B^0_{phys}(t)\ra\ell^-\nu X)+
\Gamma(\bar B^0_{phys}(t)\ra\ell^+\nu X)}=
{|q/p|^4-1\over|q/p|^4+1},\eqn\asldef$$
can be made on the basis of quark diagrams calculation of $\Gamma_{12}$
(see refs. $\(\bigi,\ssi\)$ and references therein):
$$\eqalign{a_{SL}(B^0)&\approx{8\pi\over f_2(y_t)}{m_c^2\over m_t^2}
{J\over|V_{tb}V_{td}^*|^2}\sim10^{-3},\cr
a_{SL}(B_s)&\approx{8\pi\over f_2(y_t)}{m_c^2\over m_t^2}
{J\over|V_{tb}V_{ts}^*|^2}\sim10^{-4},\cr}\eqn\aslsm$$
($J$ is the Jarlskog measure of CP violation). The estimates \aslsm\
have hadronic uncertainties of a factor of 2--3. In addition,
the estimate of $a_{SL}(B^0)$ has a large uncertainty from
the poorly determined CKM parameter $|V_{td}|$.
Again, a very large leptonic asymmetry, say $\gsim10^{-2}$, would be
difficult to explain by hadronic uncertainties and would imply
new physics.

The cleanliness of CP violation in the interference of mixing and decay
makes it the prime candidate for discovery of New Physics.
The Standard Model predictions for various classes of asymmetries
are given in Tables 2 and 3. (The signs of the asymmetries in the
last column corresponds to CP even final hadronic states and not
necessarily for the actual example in the first column.)
\vskip 1cm
\begintable
\multispan{3}\tstrut \ 2. CP Asymmetries in $B^0$ Decays \crthick
Final | Quark | SM \nr
State | Sub-Process | Prediction \crthick
$\psi K_S$|$\bar b\ra\bar cc\bar s$|$-\sin2\beta$\cr
$D^+D^-$|$\bar b\ra\bar cc\bar d$|$-\sin2\beta$\cr
$\pi^+\pi^-$|$\bar b\ra\bar uu\bar d$|$\sin2\alpha$\cr
$\phi K_S$|$\bar b\ra\bar ss\bar s$|$-\sin2(\beta-\beta^\prime)$\cr
$K_S K_S$|$\bar b\ra\bar ss\bar d$|$0$\endtable
\vskip 1cm
\begintable
\multispan{3}\tstrut \ 3. CP Asymmetries in $B_s$ Decays \crthick
Final | Quark | SM \nr
State | Sub-Process | Prediction \crthick
$\psi\phi$|$\bar b\ra\bar cc\bar s$|$-\sin2\beta^\prime$\cr
$\psi K_S$|$\bar b\ra\bar cc\bar d$|$-\sin2\beta^\prime$\cr
$\rho K_S$|$\bar b\ra\bar uu\bar d$|$-\sin2(\gamma+\beta^\prime)$\cr
$\phi\phi$|$\bar b\ra\bar ss\bar s$|$0$\cr
$\phi K_S$|$\bar b\ra\bar ss\bar d$|$\sin2(\beta-\beta^\prime)$\endtable

The various angles that appear in Tables 2 and 3 are defined by
$$\eqalign{
\alpha=\arg\left\(-{V_{td}V_{tb}^*\over V_{ud}V_{ub}^*}\right\),\ \ \ &
\gamma=\arg\left\(-{V_{ud}V_{ub}^*\over V_{cd}V_{cb}^*}\right\),\cr
\beta=\arg\left\(-{V_{cd}V_{cb}^*\over V_{td}V_{tb}^*}\right\),\ \ \ &
\beta^\prime=\arg\left\(-{V_{cs}V_{cb}^*\over V_{ts}V_{tb}^*}\right\).
\cr}\eqn\defangles$$
Of these angles, $\beta^\prime$ is constrained to be very small,
$$|\sin2\beta^\prime|\leq0.06.\eqn\smbp$$
The Standard Model constraints on $\sin2\alpha$ and $\sin2\beta$
are given in Fig. 1. (We focus on these two angles because they
are likely to be measured first.)

\FIG\figA{The Standard Model predictions in the
$\sin2\alpha$(horizontal) -- $\sin2\beta$(vertical) plane
for $110\leq m_t\leq 180\ GeV$. (The allowed ranges for all
other parameters are taken from $\(\nisa\)$.}

\topinsert
   \tenpoint \baselineskip=12pt   \narrower
 \plotpicture{\hsize}{3in}{bsga.topdraw}
\vskip12pt\noindent
{\bf Fig.~\figA.}\enskip
 The Standard Model predictions in the
$\sin2\alpha$(horizontal) -- $\sin2\beta$(vertical) plane
for $110\leq m_t\leq 180\ GeV$. (The allowed ranges for all
other parameters are taken from $\(\nisa\)$.)
\endinsert

It follows that there are several clean signals of new physics:
\item{(i)} $a_{CP}(B\ra\psi K_S)$ that is significantly smaller
than +0.2 (and certainly if it is negative).
\item{(ii)} $a_{CP}(B\ra\psi K_S)$ and $a_{CP}(B\ra\pi\pi)$
both significantly smaller than +0.5.
\item{(iii)} Any of $a_{CP}(B_s\ra\psi\phi)$,
$a_{CP}(B_s\ra\psi K_S)$ and $a_{CP}(B_s\ra\phi\phi)$
above a few percent in absolute value.
\par
\vskip 1cm

\leftline{\bf 3.\ \ \ BEYOND THE STANDARD MODEL - GENERAL}
CP asymmetries in $B$ decays are a sensitive probe of new physics
in the quark sector, because they are likely to differ from the
Standard Model predictions if there are sources of CP violation
beyond the CKM phase of the Standard Model. This can contribute
in two ways:
\par 1. If there are significant contributions to $B-\bar B$
mixing (or $B_s-\bar B_s$ mixing) beyond the box diagram
with intermediate top quarks; or
\par 2. If the unitarity of the three-generation CKM matrix
does not hold, namely if there are additional quarks.
\par Actually, there is a third way in which the Standard Model
predictions may be modified even if there are no new sources
of CP violation:
\par 3. The constraints on the CKM parameters may change if there are
significant new contributions to $B-\bar B$ mixing and to $\epsilon_K$.
\par
On the other hand, the following ingredients of the analysis of
CP asymmetries in neutral $B$ decays are likely to hold in most
extensions of the Standard Model:
\par 4. $\Gamma_{12}\ll M_{12}$. In order for this relation to
be violated, one needs a new dominant contribution to tree decays
of $B$ mesons, which is extremely unlikely, or strong suppression
of the mixing compared to the Standard Model box diagram, which
is unlikely (though not impossible for the $B_s$ system).
The argument is particularly solid for the $B_d$ system as it is
supported by experimental evidence: $\Delta M/\Gamma\sim0.7$, while
branching ratios into states that contribute to $\Gamma_{12}$ are
$\leq10^{-3}$.
\par 5. The relevant decay processes (for tree decays) are dominated by
Standard Model diagrams. Again, it is unlikely that new physics, which
typically takes place at a high energy scale, would compete with weak
tree decays. (On the other hand, for penguin dominated decays, there
could be significant contributions from new physics.)
\par
Within the Standard Model, both $B$ decays and $B-\bar B$ mixing are
determined by combinations of CKM elements. The asymmetries then
measure the relative phase between these combinations. Unitarity of
the CKM matrix directly relates these phases (and consequently
the measured asymmetries) to angles of the unitarity triangles.
In models  with new physics, unitarity of the three-generation
charged-current mixing matrix may be lost and consequently the
relation between the CKM phases and angles of the unitarity triangle
violated. But this is not the main reason that the predictions for the
asymmetries are modified. The reason is rather that if $B-\bar B$
mixing has significant contributions from new physics, the asymmetries
measure different quantities: the relative phases between the CKM
elements that determine $B$ decays and the elements of mixing matrices
in sectors of new physics (squarks, multi-scalar, etc) that contribute
to $B-\bar B$ mixing.

Thus, when studying CP asymmetries in models of new physics, we look
for violation of the unitarity constraints and, even more importantly,
for contributions to $B-\bar B$ mixing that are different
in phase and not much smaller in magnitude than the Standard Model
contribution. This leads to the following general
description of the potential for large effects in
various directions of new physics:
\par 1. In extensions of the quark sector, CKM--unitarity is violated
and there are new contributions to $B-\bar B$ mixing. Potentially,
large effects are possible.
\par 2. In Supersymmetry,
there are new contributions to $B-\bar B$ mixing. Potentially,
large effects are possible. (Note, however, that in the minimal
SUSY Standard Model (MSSM), FCNC and new phases are ``switched-off"
by hand, and no new effects are possible.)
\par 3. In extensions of the scalar sector,
there are new contributions to $B-\bar B$ mixing. Potentially,
large effects are possible. (Note, however, that in the two Higgs
doublet Model with NFC, there are no new phases,
and no new effects are possible.)
\par 4. In extensions of the gauge sector, the new gauge bosons
couple universally in flavor space. Typically, the strong constraints
from $K$-physics imply that it is unlikely to have observable
effects in $B$-physics.

In what follows, we describe several specific examples of
extensions of the Standard Model that affect CP asymmetries
in $B$ decays. The following models were discussed in detail
in the literature: 4th generation quarks
$\(\biwa,\tss,\wht,\hasu,\waka,\lond\)$; $Z$-mediated FCNC
$\(\nisib,\silv,\bpmr\)$, Left-Right Symmetry $\(\ecgr,\lowy\)$;
extensions of the scalar sector
$\(\liwo,\dogo,\gena,\sowoa,\grni,\sowo,\hawe\)$; Supersymmetry
$\(\biga,\lnsb\)$; schemes of quark mass matrices
$\(\nisa,\hara\)$; modifications of the CKM constraints
$\(\niqub,\grni\)$.
Effects of new physics on direct CP violation have been studied
in refs. $\(\tani,\thss\)$ and on CP violation in mixing in refs.
$\(\coma,\bpmra\)$.
\vskip 1cm

\leftline{\bf 4.\ \ \ SPECIFIC EXAMPLES}
\vskip 0.6cm
\leftline{\it 4.1\ \ Extra Quark Singlets $\(\nisib,\silv,\bpmr\)$}
We describe here an extension of the quark sector with an
$SU(2)_L$-singlet of charge $-1/3$. (This represents well the case when
there is such an additional quark for each generation, as in $E_6$
models.) With this extension, all the ingredients relevant to
CP asymmetries in $B$ decays are indeed affected by new physics.

In such models, the charged current mixing matrix $V$ is $3\times4$
and, most important, it is not unitary. (It is a submatrix of the
unitary $4\times4$ matrix that relates the down mass eigenstates
to the interaction eigenstates.) This leads to non-diagonal $Z$
couplings, as the neutral current mixing matrix, $U=V^\dagger V\neq1$.
In particular,
$$U_{db}=V_{ud}^*V_{ub}+V_{cd}^*V_{cb}+V_{td}^*V_{tb}\neq0.\eqn\zdb$$
Eq. \zdb\ shows that the two ingredients relevant to CP asymmetries
in $B$ decays are indeed modified in this extension:
\par 1. Unitarity of the CKM matrix is violated. In particular,
the unitarity triangle turns into a unitarity quadrangle, with
$U_{db}$ being the fourth side.
\par 2. There are new contributions to $B-\bar B$ mixing
from $Z$ mediated tree diagrams:
$$M_{12}^Z={\sqrt2\over12}G_F(B_Bf_B^2)m_B\eta(U_{db})^2.\eqn\bbz$$
\par 3. There are new sources of CP violation, as the matrices $V$
and $U$ depend on three CP violating phases.
\par
It is a peculiar property of this model that all
three new ingredients are related to each other. Let us define
the following new two angles in the unitarity quadrangle:
$$\bar\alpha=\arg\left({V_{ud}V_{ub}^*\over U_{db}^*}\right),\ \ \
\bar\beta=\arg\left({U_{db}^*\over V_{cd}V_{cb}^*}\right).\eqn\barab$$
Then, if the $Z$-mediated tree diagrams dominate $B-\bar B$ mixing,
$$a_{CP}(B\ra\psi K_S)\approx\sin2\bar\beta,\ \ \
a_{CP}(B\ra\pi\pi)\approx\sin2\bar\alpha.\eqn\newacp$$
The significant modification is then not in the new range for
$\alpha$ and $\beta$ but rather that the asymmetries now depend
on new phases, $\bar\alpha$ and $\bar\beta$. As there are no
experimental constraints on the values of $\bar\alpha$ and
$\bar\beta$ (but only on the magnitude $|U_{db}|$),
the asymmetries in \newacp\ could have any value $\(\nisib\)$,
unlike the Standard Model case described in Fig. 1.
(If the extra singlet quarks are much heavier than a few TeVs, $|U_{db}|$
is expected to be very small, the $Z$-mediated FCNC contribute
negligibly to $B-\bar B$ mixing, and the deviations from the
Standard Model predictions are unobservably small.)

In ref. $\(\silv\)$ it was shown that the upper bound on $|U_{sb}|$
from the UA1 measurement of $b\ra s\mu^+\mu^-$ implies that the
effects on CP asymmetries in $B_s$ decays cannot be maximal. For
example, the zero asymmetries predicted for various $B_s$ decays
(see Table 3), could be modified to, at most, ${\cal O}(0.3)$.
In ref. $\(\bpmr\)$ it was observed that even if the $Z$ contributions
do not dominate the mixing but are just not much smaller than the box
diagrams, they could still have large effects on the asymmetries.
In this case, the asymmetries in \newacp\ would have a more
complicated dependence on $\alpha$, $\beta$, $\bar\alpha$ and
$\bar\beta$.
\vskip 0.6cm

\leftline{\it 4.2\ \ Quark-Squark Alignment $\(\nise,\lnsb\)$}
We describe here a supersymmetric extension of the Standard Model
that is different from the MSSM. In particular, the mechanism
that suppresses SUSY-induced FCNC is not squark degeneracy.
Instead, the quark mass matrices and the squark mass-squared matrices
are naturally aligned in models of abelian horizontal
symmetry $\(\nise\)$, namely the are both approximately diagonal
in the same basis. If this alignment is precise enough, the
mixing matrix for quark-squark-gluino couplings is very close to the
unit matrix, and FCNC are highly suppressed even if squarks are
not degenerate at all.
\par
The motivation for this extension $\(\lns\)$
was to explain the hierarchy in the quark sector parameters,
$$\eqalign{1\sim&\ \ m_t/\VEV{\phi_u};\cr
\lambda\sim&\ \ V_{us};\cr
\lambda^2\sim&\ \ V_{cb},\ m_d/m_s,\ m_s/m_b;\cr
\lambda^3\sim&\ \ V_{ub},\ m_u/m_c,\ m_c/m_t.\cr}\eqn\hierarchy$$
(with $\lambda\sim0.2$ these relations hold to within a factor of 2.)
These relations are predicted and the alignment of quarks and squarks
is precise enough to satisfy the constraints from neutral meson
mixing if the mass matrices have the following form (for details
see $\(\lnsb\)$):
$$M^d\sim\VEV{\phi_d}\pmatrix{\lambda^4&0&\lambda^3\cr
0&\lambda^2&\lambda^2\cr 0&0&1\cr},\ \ \
M^u\sim\VEV{\phi_u}\pmatrix{\lambda^6&\lambda^4&\lambda^3\cr
\lambda^5&\lambda^3&\lambda^2\cr \lambda^3&\lambda^2&1\cr}.\eqn\mumd$$
(All entries here are just order of magnitude estimates.)

Such a structure for the quark mass matrices can be a result of
a horizontal (discrete subgroup of) $U(1)_a\times U(1)_b$ symmetry,
that is spontaneously broken by the VEVs of two Standard Model
singlet scalars:
$$S_a(-1,0): \ \ {\VEV{S_a}\over M}\sim\lambda;\ \ \
S_b(0,-1): \ \ {\VEV{S_b}\over M}\sim\lambda^2.\eqn\svevs$$
$M$ is a high scale where the information about the horizontal
symmetry breaking is communicated to the light quarks. An example of
charge assignments that lead to $M^d$ as in \mumd\ is the following:
$$\eqalign{Q_1(3,0),\ \ Q_2(0,1),&\ \ Q_3(0,0);\cr
\bar d_1(-1,1),\ \ \bar d_2(2,-1),&\ \ \bar d_3(0,0).\cr}\eqn\qcharge$$
Here, the $Q_i$ are quark-doublet supermultiplets, while $\bar d_i$
are down-quark singlet supermultiplets. The charge assignments
in \qcharge\ determine the form of the squark mass-squared matrices
as well. Most important for our study are the diagonal
blocks in the down-squark mass-squared matrix:
$$\tilde M^{d2}_{LL}\sim\tilde m^2\pmatrix{1&\lambda^5&\lambda^3\cr
\lambda^5&1&\lambda^2\cr \lambda^3&\lambda^2&1\cr},\ \ \
\tilde M^{d2}_{RR}\sim\tilde m^2\pmatrix{1&\lambda^7&\lambda^3\cr
\lambda^7&1&\lambda^4\cr \lambda^3&\lambda^4&1\cr}.\eqn\sqmass$$

The structure of $M^d$ and $\tilde M^{d2}$
allows an estimate of the mixing matrix for quark-squark-gluino
interaction which, in turn, gives an estimate of the SUSY
contribution to neutral meson mixing. With the mass matrices
of eqs. \mumd\ and \sqmass, SUSY contribution to $B-\bar B$ mixing
(with $\tilde m\sim m_{\tilde g}\sim1\ TeV$)
is about 20\% of the Standard Model one.
On the other hand, the SUSY contribution to mixing in the $K$ system
is negligibly small. Actually, it is small enough to obey the more
stringent $\epsilon_K$ constraints even for phases of order 1.

As the SUSY diagram is, in magnitude, about 20\% of $M_{12}(B^0)$
but with a phase that could be very different from the Standard Model
one, the Standard Model predictions for CP asymmetries in
$B^0$ decays may be modified by as much as 0.4, a sizable effect.
On the other hand, a similar analysis for $B_s$ mixing shows that
it cannot be significantly affected by the SUSY contributions,
so that the Standard Model predictions for CP asymmetries in $B_s$
decays remain unchanged.

The quark-squark alignment mechanism has strong testable
predictions, namely that squarks are not degenerate and that
$D-\bar D$ mixing is close to the experimental upper bound.
Large effects on CP asymmetries in $B$ decays are not a necessary
result of quark-squark alignment, but their measurement would be
extremely useful in distinguishing between various explicit models that
incorporate this mechanism. Furthermore, the model above shows that
the absence of modifications to the Standard Model predictions for
CP asymmetries in $B$ decays in the MSSM is a special property of
this model and not a generic feature of SUSY models.
\vskip 0.6cm

\leftline{\it 4.3\ \ Charged Scalar Exchange $\(\grni\)$}
In models of three or more scalar doublets, the mixing matrix for
charged scalars contains one or more
CP violating phases. This phase could,
in principle, affect CP asymmetries in $B$ decays $\(\grni\)$.
However, recent experimental constraints imply that the effect is
too small to be observed. Still, the Standard Model predictions
may be violated because the constraints on the CKM parameters change.

In multi-scalar models, $B-\bar B$ mixing gets additional contributions
from box diagrams where one or two of the Standard Model $W$-boson
propagators are replaced by the charged scalar $H$ propagators.
This situation can be presented in the following way:
$$M_{12}(B^0)={G_F^2\over 64\pi^2}(V_{td}^*V_{tb})^2(I_{WW}+2I_{WH}
+I_{HH}),\eqn\mixcse$$
where $I_{WW}$, $I_{HW}$ and $I_{HH}$ are functions of
the intermediate particle masses ($m_W$, $m_H$ and $m_t$) and of the
Yukawa couplings. The Standard Model contribution is $I_{WW}$.
The functions $I_{HW}$ and, in a more significant way, $I_{HH}$
depend on the phase in the charged scalar mixing matrix.

Let us define a phase $\theta_H$ according to
$$\theta_H=\arg(I_{WW}+2I_{WH}+I_{HH}).\eqn\thetah$$
($I_{WW}$ is real, so that in the Standard Model $\theta_H=0$.)
The angles measured by CP asymmetries in $B^0$ decays will be
universally shifted by $\theta_H$. Specifically,
$$a_{CP}(B\ra\psi K_S)=-\sin(-2\beta+\theta_H),\ \ \
a_{CP}(B\ra\pi\pi)=\sin(2\alpha+\theta_H).\eqn\cseacp$$

The magnitude of this effect depends on how large $\theta_H$ is.
Existing constraints from CP violating processes, most noticeably
the electric dipole moment of the neutron, still allow for
very large $\theta_H$. However, the CP violating charged scalar
couplings contribute also to the CP conserving decay $b\ra s\gamma$.
The recent CLEO bound on the rate of this decay gives the strongest
constraint on CP violation from charged scalar exchange $\(\grni\)$.
It implies
that the effect on CP asymmetries in $B^0$ decays cannot be larger
than 2\%, too small to stand out as a signal of new physics.

Modifications of the Standard Model predictions for CP asymmetries
in $B$ decays may also arise from the different constraints on
CKM parameters. This holds even for two scalar doublet (type I and
type II) models where indeed there are no new phases. The most
significant effect is that the lower bounds on $|V_{tb}V_{td}^*|$
from $B-\bar B$ mixing and from $\epsilon_K$ are relaxed, because
charged scalar exchange may contribute significantly. This situation is
actually much more general than our specific multi-scalar framework,
and the results below apply to all models with significant contributions
to $x_d$ and $\epsilon_K$: a new region (forbidden in the Standard Model)
opens up in the plane of $\sin2\alpha-\sin2\beta$,
as shown in Fig. 2 $\(\grni\)$. If experiment finds a relatively low
value of $\sin2\beta$ (below 0.5) and a negative value for
$\sin2\alpha$, it may be an indication that there are significant
contributions from new physics to $B-\bar B$ mixing, even if these
contributions carry no new phases.

\FIG\figB{The allowed region in the $\sin2\alpha$ -- $\sin2\beta$ plane
in the Standard Model (solid) and the new allowed region
in multi-scalar models (dot-dashed).}

\topinsert
   \tenpoint \baselineskip=12pt   \narrower
 \plotpicture{\hsize}{3in}{bsga.topdraw}
\vskip12pt\noindent
{\bf Fig.~\figB.}\enskip
The allowed region in the $\sin2\alpha$ -- $\sin2\beta$ plane
in the Standard Model (solid) and the new allowed region
in multi-scalar models (dot-dashed).
\endinsert

Multi-scalar models without NFC are much less constrained,
and may give large effects on the CP asymmetries $\(\sowo\)$.
An interesting case is that of light scalars with small couplings to
quarks protected by approximate symmetries, where close to zero
asymmetries are expected for all $B$ decays $\(\hawe\)$.
\vskip 0.6cm

\leftline{\it 4.4\ \ Schemes for Quark Mass Matrices $\(\nisa\)$}
As far as CP asymmetries in $B$ decays are concerned,
extensions of the Standard Model that provide relations between
the quark sector parameters are unique: instead of relaxing the
Standard Model constraints on CP asymmetries in $B$ decays,
they actually narrow down considerably the allowed ranges.
This means that while none of the extensions discussed in
previous sections can be excluded on the basis of measurements
of CP asymmetries, schemes for quark mass matrices can.

We will not go to any details concerning the various schemes
for quark mass matrices discussed here. Instead, we present
in Fig. 3 $\(\nisa\)$ the predictions for $a_{CP}(B\ra\psi K_S)$ and
$a_{CP}(B\ra\pi\pi)$ from schemes by Fritzsch (the thin black wedge
in Fig. 3.a); Giudice (the black band in Fig. 3.b);
Dimopoulos-Hall-Raby (the black region in Fig. 3d);
and the ``symmetric - CKM" scheme (the black curves in Figs. 3.c and
3.d). (For detailed references, see $\(\nisa\)$.)
It is clear from the figure that CP asymmetries in the
above-mentioned modes would crucially test each of these schemes.

\FIG\figC{The regions predicted by various mass matrix schemes in the
 $\sin2\alpha$ -- $\sin2\beta$ plane for $m_t=$ (a) 90 GeV,
(b) 130 GeV, (c) 160 GeV, (d) 185 GeV. The Standard Model predictions
are outlined in grey, and those of the various schemes in black.
(See the text for details.)}

\topinsert
   \tenpoint \baselineskip=12pt   \narrower
 \plotpicture{\hsize}{4in}{bsga.topdraw}
\vskip12pt\noindent
{\bf Fig.~\figC.}\enskip
The regions predicted by various mass matrix schemes in the
 $\sin2\alpha$ -- $\sin2\beta$ plane for $m_t=$ (a) 90 GeV,
(b) 130 GeV, (c) 160 GeV, (d) 185 GeV. The Standard Model predictions
are outlined in grey, and those of the various schemes in black.
(See the text for details.)
\endinsert

\vskip 1cm

\leftline{\bf 5.\ \ \
HOW TO DISTINGUISH BETWEEN VARIOUS TYPES}
\leftline{\bf \ \ \ \ \ OF NEW PHYSICS?}
If deviations from the Standard Model predictions
are found, how can we tell which extension of the Standard Model
(among the many extensions that allow large effects) is responsible
for that? In this chapter, we show that the richness of experimental
measurements, reflected in the large number of modes in Tables 2
and 3, can be used to study very detailed features of the new
physics that might affect the CP asymmetries $\(\nisia,\niqub\)$.

More specifically, various relations among the asymmetries
do not depend on all the assumptions that go into the analysis
and thus may hold beyond the Standard Model or, conversely,
if they are violated can help pinpoint which ingredients
must be added to the Standard Model.
Here are a few examples.
\par (i) Violation of
$$a_{CP}(B\ra D^+D^-)=-a_{CP}(B\ra\psi K_S)\eqn\kktest$$
(the minus sign comes from the opposite CP of the final states)
would imply that (a) there is new physics contribution to
$K-\bar K$ mixing and (b) the approximate unitarity relation
$V_{ud}^*V_{us}+V_{cd}^*V_{cs}\approx0$ (where we neglected
$V_{td}^*V_{ts}$) is violated.
\par (ii) Violation of
$$a_{CP}(B_s\ra\psi\phi)\approx0\eqn\bstest$$
would imply that there is new physics contribution to $B_s-\bar B_s$
mixing. As shown in ref. $\(\nisia\)$, this condition is equivalent to
$$\alpha+\beta+\gamma=\pi\eqn\bstesta$$
(where $\alpha$, $\beta$ and $\gamma$ are deduced from the CP asymmetries
in $B\ra\pi\pi$, $B\ra\psi K_S$ and $B_s\ra\rho K_S$, respectively).
\par (iii) Violation of
$$a_{CP}(B\ra\psi K_S)=\sin2\beta,\ \
a_{CP}(B\ra\pi\pi)=\sin2\alpha,\eqn\bstest$$
(where $\sin2\alpha$ and $\sin2\beta$ are calculated from the
constraints on the unitarity triangle)
would imply that there is new physics contribution to $B^0-\bar B^0$
mixing.
\par (iv) Violation of
$$a_{CP}(B_s\ra\psi\phi)\approx a_{CP}(B_s\ra\phi\phi)\eqn\bsunitest$$
would most likely imply that the approximate unitarity relation
$V_{cb}^*V_{cs}+V_{tb}^*V_{ts}\approx0$ (where we neglected
$V_{ub}^*V_{us}$) is violated.

As an example, we explain the test (i) above.
The phases measured by the two modes are:
$$\eqalign{
\arg\lambda(B\ra D^+D^-)=&
\arg(M_{12}(B^0))-2\arg(A(\bar b\ra\bar cc\bar d)),\cr
\arg\lambda(B\ra\psi K_S)=&
\arg(M_{12}(B^0))-2\arg(A(\bar b\ra\bar cc\bar s))-\arg(M_{12}(K^0)).\cr
}\eqn\kktesta$$
It is clear that the phase of the $B^0$ mixing amplitude does
not affect the relation of eq. \kktest\ (even though it affects the
actual values of the asymmetries). As decay amplitudes are dominated by
$W$-mediated tree diagrams, \kktest\ does hold if
$$\arg(M_{12}(K^0))=\arg((V_{cd}V_{cs}^*)^2).\eqn\kktestb$$
This is trivially the case if $K-\bar K$ mixing is dominated
by the Standard Model box diagram with virtual $c$ quarks.
Therefore, a necessary condition for violating \kktest\ is
a new mechanism for $K-\bar K$ mixing. However, the extremely
small experimental value of $\epsilon_K$ implies that
$\arg(M_{12}(K))/\Gamma_{12}(K))\sim10^{-3}$. Therefore,
model-independently
$$\arg(M_{12}(K^0))\approx\arg((V_{ud}V_{us}^*)^2).\eqn\kktestc$$
Consequently, another necessary condition for violating \kktest\ is that
$V_{ud}V_{us}^*+V_{cd}V_{cs}^*\neq0$.

We conclude that with CP asymmetries measured in many $B$ decay
modes, we can learn many detailed features of the new physics
that affects their values.

\vskip 1cm

\leftline{\bf ACKNOWLEDGEMENTS}
This work is based on collaborations with Claudio Dib, Yuval Grossman,
Miriam Leurer, David London, Helen Quinn, Uri Sarid, Nati Seiberg
and Dennis Silverman. YN is an incumbent of the Ruth E. Recu Career
Development Chair, and is supported in part by the Israel Commission
for Basic Research, by the United States -- Israel Binational
Science Foundation (BSF), and by the Minerva Foundation.
\vskip 1cm
\refout
\end